\documentclass[pra,aps,twocolumn]{revtex4}

\usepackage{amsmath}
\usepackage{epsfig}

\begin{document}
\title{Range criterion and classification of true entanglement in $2\times{M}\times{N}$ system}
\author{Lin Chen}
\author{Yi-Xin Chen}
\affiliation{Zhejiang Insitute of Modern Physics, Zhejiang
University, Hangzhou 310027, People's Republic of China}

\begin{abstract}
We propose a range criterion which is a sufficient and necessary
condition satisfied by two pure states transformable with each
other under reversible stochastic local operations assisted with
classical communication. We also provide a systematic method for
seeking all kinds of true entangled states in the
$2\times{M}\times{N}$ system, and can effectively distinguish them
by means of the range criterion. The efficiency of the criterion
and the method is exhibited by the classification of true
entanglement in some types of the tripartite systems.
\end{abstract}
\maketitle

One of the main tasks in quantum information theory (QIT) is to
find out how many different ways there exist, in which several
spatially distributed objects could be entangled under certain
prior constraints of physical resource. Such a restriction is the
transformation of entangled states by local operations assisted
with classical communication (LOCC). All bipartite pure entangled
states are interconvertible in the asymptotic LOCC transformations
\cite{Bennett1}.  This implies that all bipartite pure entangled
states can be used to perform the same task of entanglement
processing in the asymptotic regime. However, for single copies of
states, any two pure states in the same class under LOCC can be
convertible with certainty from each other by local unitary
operation \cite{Vidal,Bennett2}.

In the general bipartite system, infinitely many entangled states
are not related by local unitary operations and continuous
parameters are used to characterize all equivalent classes
corresponding to them \cite{Linden,Kempe}. In order to classify
them more succinctly, an alternative restriction was introduced
\cite{Bennett2,Dur}. This restriction requires that the conversion
of the states is through stochastic local operations and classical
communication (SLOCC), i.e. through LOCC but without imposing that
it has to be achieved with certainty. Two pure states of a
multipartite system are equivalent under SLOCC if and only if
(iff) they are related by an invertible local operator (ILO)
\cite{Dur}. Comparing with the other constraints, SLOCC is
coarser, but more practical and simpler for the classification of
entanglement. For instance, the state $\left|00\right\rangle$ and
the Bell state, $\left|\Phi\right\rangle
=\left|00\right\rangle+\left|11\right\rangle$ are under this
criterion the only two classes in two-qubit systems.

Such restriction becomes stricter in multipartite settings, since
multiparty entanglement has a much more complicated configuration
than the bipartite case. For the true tripartite entangled states,
there exist not only the three-qubit GHZ state \cite{Greenberger}
and but also the so-called three-qubit W state \cite{Dur}, which
was shown to be essentially different from the GHZ state under
SLOCC. In principle, the conclusion of the SLOCC equivalence under
ILO's \cite{Dur} is sufficient to classify the entanglement
properties of single copies of states. However, for the
classification of entangled states of general multipartite system,
the classification based on this conclusion becomes more and more
complicated with increasing of dimensions of Hilbert space of such
system and the exponential increase of number of parameters when
classifying multiparty entanglement
\cite{Verstraete1,Miyake2,Miyake3}. In fact, it is almost
impossible to make use of it to perform the classification and the
construction of the different entangled states in general
multipartite system. So we need to establish a more effective
criterion involved in the subspaces of the Hilbert space.

In this paper, we introduce a more straightforward and effective
criterion of the equivalent classes of entanglements under SLOCC
in general multipartite system, where only a single copy of state
is available. Different from the definition of equivalence classes
of entangled states suggested by D\"ur \textit{et al } \cite{Dur},
our criterion is constructive for the entangled states of
multipartite system. Indeed, based on the criterion, an iterated
method can be introduced to determine all classes of true
entangled states in the $2\times M\times N$ system. Combined such
method with the criterion, we not only can classify the entangled
states but also can construct them.

In order to arrive at our criterion in a smooth way, we ought to
introduce some necessary notations and useful concepts in this
paper. Since we are concerned with the entanglement properties of
the states of many parties, we should understand the quantum
properties of the states of each party. Such quantum state can be
described by the reduced density matrix of single party from the
composite state of many parties. The rank of it is called as the
local rank. Since we are interested with the classification of
pure entangled states in the multipartite system, the state
$\rho_{A_1 A_2 \cdots A_N}$ can be realized by the expression
$\rho_{A_1 A_2 \cdots
A_N}=\left|\Psi\right\rangle_{{A_1}{A_2}\cdots
{A_N}}\left\langle\Psi\right|$. In the notation, we shall use
$\rho^B_{\Psi_{AB}}$ to stand for the reduced density operator of
$B$ system from the state $\left|\Psi\right\rangle_{AB}$. The
local ranks of any pure state are invariant under SLOCC
\cite{Dur}. Hence, one can use the local ranks of the parties to
characterize the Hilbert space of multipartite system, e.g., the
$D_1\times D_2\times \cdots\times D_N$ space, where $D_i$ is the
local rank of the party $A_i$. Due to the invariance of the local
ranks under SLOCC and the Schmidt decomposition with respect to
the party $A_j$ and the other party $A_1 A_2\cdots A_{j-1}
A_{j+1}\cdots A_N$, any state
$\left|\Psi\right\rangle_{{A_1}{A_2}\cdots{A_N}}$ ( sometimes also
written $\left|\Psi\right\rangle_{{D_1}{D_2}\cdots{D_N}}$ ) in the
$D_1\times D_2\times\cdots\times D_N$ space can always be
transformed into the following form,
\begin{equation}
\left|\Phi\right\rangle=\sum_{i=0}^{D_j-1}
\left|i\right\rangle_{A_j}\otimes\left|i\right\rangle_{{A_1}{A_2}\cdots{A_{j-1}}{A_{j+1}}\cdots{A_N}},
\end{equation}
where $\langle{i}|{k}\rangle_{A_j}={\delta}_{ik}$ and
$\{\left|i\right\rangle_{{A_1}{A_2}\cdots{A_{j-1}}{A_{j+1}}\cdots{A_N}},i=0,1,\cdots,D_j-1\}$
are a set of linearly independent vectors. We call the above
expression $adjoint$ $form$ and the vector
$\left|i\right\rangle_{{A_1}{A_2}\cdots{A_{j-1}}{A_{j+1}}\cdots{A_N}}$
$adjoint$ $state$. Here the set $\{\left|i\right\rangle_{A_j}\}$
are chosen as the computational basis.

The concept of range of quantum state plays an essential role in
our criterion. Let state $\rho$ act on the Hilbert space ${\cal
H}$. In the standard manner, the range of $\rho$ is defined by
$R(\rho)=\left|\Psi\right\rangle\in {\cal
H}:\rho\left|\Phi\right\rangle =\left|\Psi\right\rangle$, for some
$\left|\Phi\right\rangle\in {\cal H}$. For the $adjoint$ reduced
density matrix $\rho_{\Psi_{{A_1}{A_2}...{A_N}}}^{A_1 A_2\cdots
A_{j-1} A_{j+1}\cdots A_N}$ of the party $A_j$, one can clearly
see that all states $\left|\Theta\right\rangle$, which are from
$\left|\Theta\right\rangle
=\rho_{\Psi_{{A_1}{A_2}\cdots{A_N}}}^{A_1 A_2\cdots A_{j-1}
A_{j+1}\cdots A_N}\left|\Gamma\right\rangle $ for any
$\left|\Gamma\right\rangle\in {\cal H}_{A_1 A_2\cdots A_{j-1}
A_{j+1}\cdots A_N}$, span the whole range of it. In fact, for a
general multipartite system, the local rank of each party and the
range of the adjoint reduced density matrix of it determine
completely the character property of a pure multiple entangled
state under SLOCC. The following theorem exhibits such relation.

\textit{Theorem 1: Range Criterion.} Two pure states of a
multipartite system are equivalent under SLOCC iff (i) they have
the same local rank of each party, and (ii) the ranges of the
adjoint reduced density matrices of each party of them are related
by certain ILO's.

Now, let us use $\left|\Psi\right\rangle_{{A_1}{A_2}\cdots{A_N}}$
and $\left|\Phi\right\rangle_{{A_1}{A_2}\cdots{A_N}}$ to denote
such two states, and $V_i$ to do the ILO acting on the party
$A_i$. The theorem 1 can be formulated that
$\left|\Psi\right\rangle_{{A_1}{A_2}\cdots{A_N}}=
{V_1}\otimes{V_2}\otimes\cdots\otimes{V_N}\left|\Phi\right\rangle_{{A_1}{A_2}\cdots{A_N}}$
iff they have the same local rank, $S_1\subseteq S_2$ and
$S_2\subseteq S_1$, i.e. $S_1 =S_2$, where
\begin{eqnarray}
S_1&\equiv&\{\left|\phi\right\rangle_{{A_2}\cdots{A_N}},\left|\phi\right\rangle\in
R(\rho^{{A_2}\cdots{A_N}}_{\Psi_{{A_1}{A_2}\cdots{A_N}}})\},\\
S_2&\equiv&\{\left|\phi\right\rangle_{{A_2}\cdots{A_N}},\left|\phi\right\rangle\in
{V_2}\otimes...\otimes{V_N}R(\rho^{{A_2}\cdots{A_N}}_{\Phi_{{A_1}{A_2}\cdots{A_N}}})\}\nonumber.
\end{eqnarray}

\textit{Proof}. \textit{Necessity}. If the two states are
equivalent under SLOCC, they must have the same local rank. By
means of the expression of
$\left|\Phi\right\rangle_{{A_1}{A_2}\cdots{A_N}}$ in the adjoint
form, we get $
R(\rho^{{A_2}\cdots{A_N}}_{\Psi_{{A_1}{A_2}\cdots{A_N}}})=\sum_{i,j=0}^{D_1-1}
\left\langle{i}\right|{V^\dagger_1}V_1\left|{j}\right\rangle_{A_1}
\left\langle{i}\right|{V^\dagger_2}\otimes\cdots\otimes
{V^\dagger_N}\left|{\mu}\right\rangle_{{A_2}\cdots{A_N}}{V_2}
\otimes\cdots\otimes{V_N}\left|{j}\right\rangle_{{A_2}\cdots{A_N}}
\equiv\sum_{j=0}^{D_1-1}
{C_j}({V_2}\otimes\cdots\otimes{V_N}\left|{j}\right\rangle_{{A_2}\cdots{A_N}})$,
and $R(\rho^{{A_2}\cdots{A_N}}_{\Phi_{{A_1}{A_2}\cdots{A_N}}})=
\sum_{j=0}^{D_1-1}\langle{j}|{\omega}\rangle_{{A_2}\cdots{A_N}}\left|{j}\right\rangle_{{A_2}\cdots{A_N}}.
$ Here, $\left|{\mu}\right\rangle$ and
$\left|{\omega}\right\rangle$ are two arbitrary vectors in the
$A_2\cdots A_N$ space, and we can write
$\left|{\omega}\right\rangle=(\omega_0,\omega_1,...,\omega_{d-1})^T,
d=D_2\times D_3\times...\times D_N$. We prove that there always
exists a vector $\left|{\omega}\right\rangle$ which satisfies
$C_j=\langle{j}|{\omega}\rangle,j=0,1,...,D_1-1$, namely,
$M_{D_1\times{d}}(\omega_0,\omega_1,...,\omega_{d-1})^T=(C_0,C_1,...,C_{D_1-1})^T$,
where the coefficient matrix $M$ consists of $j$'s index entering
the row. Notice that
$\{\left|i\right\rangle_{{A_2}\cdots{A_N}},i=0,1,...,D_1-1\}$ are
linearly independent, so $D_1\leq d$. This implies the existence
of $\left|{\omega}\right\rangle$. Since $V_i$'s are invertible,
the above discussion shows $S_1\subseteq S_2$. Similarly we can
get $S_2\subseteq S_1$($\Phi\leftrightarrow\Psi$,
$V_i\leftrightarrow{V^{-1}_i}$).

\textit{Sufficiency}. Suppose
$\left|\Psi\right\rangle_{{A_1}{A_2}\cdots{A_N}}$ and
$\left|\Phi\right\rangle_{{A_1}{A_2}\cdots{A_N}}$ have the same
local rank. The above result for $S_2\subseteq S_1$ can be more
explicitly expressed as: if we write
$\left|\Phi\right\rangle=\sum_{i=0}^{D_1-1}
\left|i\right\rangle_{A_1}\otimes\left|\varphi_i\right\rangle_{{A_2}\cdots{A_N}}$
and $\left|\Psi\right\rangle=\sum_{i=0}^{D_1-1}
\left|i\right\rangle_{A_1}\otimes\left|\psi_i\right\rangle_{{A_2}\cdots{A_N}}$
, then
${V_2}\otimes\cdots\otimes{V_N}\left|{\varphi_j}\right\rangle=\sum_{i=0}^{D_1-1}a_{ji}
\left|\psi_i\right\rangle,j=0,1,...,D_1-1.$ This expression can be
equivalently written as  $V^{d\times
d}\left|j\right\rangle=\sum_{i=0}^{D_1-1}a_{ji}
\left|i\right\rangle,j=0,1,...,D_1-1,$ where the
$\left|i\right\rangle$'s are computational basis and $V^{d\times
d}$ is invertible. So the matrix $M^{D_1\times D_1}$ is invertible
where $M_{ij}=a_{ji}$. Thus, we have
$\sum_{j=0}^{D_1-1}\left|j\right\rangle_{A_1}
{V_2}\otimes\cdots\otimes{V_N}\left|{\varphi_j}\right\rangle_{{A_2}\cdots{A_N}}=
\sum_{j=0}^{D_1-1}M^T\left|j\right\rangle_{A_1}
\left|{\psi_j}\right\rangle_{{A_2}\cdots{A_N}}$. By setting
$V_1=(M^T)^{-1}$, we arrive at the conclusion of the theorem. For
the case of $S_1\subseteq S_2$, the similar result can be proved.
Q.E.D.

The above theorem gives a universal criterion of the equivalent
classes of multiparty entanglement under SLOCC. It shows that one
can judge whether two N-partite states are equivalent under SLOCC
or not by analyzing the range of the reduced density operator of
any $N-1$ parties (the ranks are easily obtained). When two given
states $\left|\Psi\right\rangle_{{A_1}{A_2}\cdots{A_N}}$ and
$\left|\Phi\right\rangle_{{A_1}{A_2}\cdots{A_N}}$ are
investigated, we have to write out those local operations in
detail, i.e., $V^{D_i\times D_i}_i=[a_{ijk}]$, the indices of
parties, $i=0,1,...,N-1,$ that of the matrix entries
$j,k=0,1,...,D_i-1$, and they map one range to another. Obviously,
whether the resulting entries $a_{ijk}$'s keep the nonsingularity
of all $V_i$'s determines whether the two states are equivalent
under SLOCC or not. To simplify the calculation, one can determine
the number of product states in the range of the adjoint reduced
density matrix of each party. Such the number of product states
can be accounted from the state expansions in the range basic
vectors to satisfy the condition of product state. For example,
consider the two stand forms of three-qubit states,
$\left|GHZ\right\rangle_{ABC}=\left|000\right\rangle+\left|111\right\rangle$
and
$\left|W\right\rangle_{ABC}=\left|001\right\rangle+\left|010\right\rangle+\left|100\right\rangle$.
One can readily write out
$R(\rho^{AB}_{GHZ})=\alpha_0\left|00\right\rangle+\alpha_1\left|11\right\rangle$
, and
$R(\rho^{AB}_{W})=\beta_0\left|00\right\rangle+\beta_1(\left|01\right\rangle+\left|10\right\rangle)$,
$\alpha_0,\alpha_1,\beta_0,\beta_1\in C.$ Hence, no matter how the
coefficients change, there are two product states
$\left|00\right\rangle$ and $\left|11\right\rangle$ in
$R(\rho^{AB}_{GHZ})$, and only one product state
$\left|00\right\rangle$ in $R(\rho^{AB}_{W})$, which has been used
for the existence of different types of entanglement in
three-qubit states \cite{Dur}. The reason is, the states
consisting of adjoint states are always different
\cite{notation1}. Since any product state can only be transformed
into another product state under ILO's, we obtain a useful
corollary as follows.

\textit{Corollary 1}. If two pure states of a multipartite system
are equivalent under SLOCC, the numbers of product states in the
ranges of the adjoint reduced density matrices of each party of
them must be equal.

The above corollary gives a necessary condition of equivalent
classes of multiparty entanglement under SLOCC, and it is by
employing this condition often more practical than the analysis of
the whole range of given states, e.g., two states
$\left|GHZ\right\rangle$ and $\left|W\right\rangle$ are
inequivalent under SLOCC, since we have known they have different
numbers of product states in the ranges. However, the equality of
the numbers of product states in the ranges of two states does not
imply the equivalence of them under SLOCC. Once that the equality
of the numbers of product states in the ranges is satisfied, one
must check whether the states in the ranges can be transformed
with each other under ILO's or not. If yes, such two states belong
to an equivalent class under SLOCC. Otherwise, they are not
equivalent under SLOCC. We introduce a notation
$[a_1,a_2,a_3,...,a_N]$ representing a set of states, and a state
$\left|\Phi\right\rangle_{{A_1}{A_2}\cdots{A_N}}\in[a_1,a_2,a_3,...,a_N]$
iff the number of product states in
$R(\rho^{{A_1}\cdots{A_{i-1}}{A_{i+1}}\cdots{A_N}}_{\Phi_{{A_1}{A_2}\cdots{A_N}}})$
is $a_i$, $i=2,...,N-1$. We will characterize the entanglement
property of the state
$\left|\Phi\right\rangle_{{A_1}{A_2}\cdots{A_N}}$ mainly by this
notation.

We examine the above depiction with a concrete example. Consider
two states in the $2\times4\times4$ system, $
\left|\phi_0\right\rangle_{ABC}=\left|000\right\rangle+\left|111\right\rangle+
\left|022\right\rangle+\left|033\right\rangle,$ and $
\left|\phi_1\right\rangle_{ABC}=\left|001\right\rangle+\left|010\right\rangle+\left|100\right\rangle+
\left|022\right\rangle+\left|033\right\rangle.$ Notice both of
them possess symmetry under exchange of particle $B$ and $C$, so
we can write
$R(\rho^{AB}_{\phi_0})=R(\rho^{AC}_{\phi_0})=\alpha_0\left|00\right\rangle
+\alpha_1\left|11\right\rangle+\alpha_2\left|02\right\rangle+\alpha_3\left|03\right\rangle$,
and $R(\rho^{BC}_{\phi_0})=\beta_0(\left|00\right\rangle
+\left|22\right\rangle+\left|33\right\rangle)+\beta_1\left|11\right\rangle,
\alpha_0,\alpha_1,\alpha_2,\alpha_3,\beta_0,\beta_1\in C$. Let us
move to find out the number of product states in all three ranges.
one can readily see there are infinitely many product states in
$R(\rho^{AB}_{\phi_0})$ (also $R(\rho^{AC}_{\phi_0})$), e.g., by
supposing $\alpha_0=\alpha_1=0$ and then changing $\alpha_2$ and
$\alpha_3$. On the other hand, supposing $\beta_0=0$ shows
$\left|11\right\rangle$ is the unique product state in
$R(\rho^{BC}_{\phi_0})$. According to the above notation, we then
write
\begin{equation*}
\left|\phi_0\right\rangle_{ABC}=\left|000\right\rangle+\left|111\right\rangle+
\left|022\right\rangle+\left|033\right\rangle\in[1,\infty,\infty].
\end{equation*}
In a similar way, we obtain
\begin{equation*}
\left|\phi_1\right\rangle_{ABC}=\left|001\right\rangle+\left|010\right\rangle+\left|100\right\rangle+
\left|022\right\rangle+\left|033\right\rangle\in[1,\infty,\infty].
\end{equation*}
Clearly, the numbers of product states in each range of
$\left|\phi_0\right\rangle$ and $\left|\phi_1\right\rangle$ are
equal. For judging whether they are equivalent under SLOCC or not,
we have to analyze the detailed structure of range of them. First,
write out
$R(\rho^{BC}_{\phi_1})=\beta_0(\left|01\right\rangle+\left|10\right\rangle
+\left|22\right\rangle+\left|33\right\rangle)+\beta_1\left|00\right\rangle$,
whose local rank of system B (or C) is 4 or 1. According to the
range criterion, it must be mapped into $R(\rho^{BC}_{\phi_0})$ by
some ILO's. Because the local rank of system B (or C) in
$R(\rho^{BC}_{\phi_0})$ is 3 or 1, there exist no ILO's making
this transformation. $\left|\phi_0\right\rangle$ and
$\left|\phi_1\right\rangle$ are thus inequivalent under SLOCC,
although they have the same number of product states in each
range.

Now, let us focus on the classification of entanglement in the
$2\times M\times N$ system. Although the structure of the whole
space of pure states of this system
$\left|\Psi\right\rangle_{2\times M\times N}$ is very complicated,
we can establish an interesting relation between
$\left|\Psi\right\rangle_{2\times M\times N}$ and some spaces of
pure states with the local ranks lower than those of it. We shall
use $\sim$ to denote the equivalence under SLOCC in this paper.
There are two useful ILO's in the below discussion. They are
defined by $O^A_1(\left|\phi\right\rangle,\alpha):
\left|\phi\right\rangle_A\rightarrow\alpha\left|\phi\right\rangle_A$
and $O^A_2(\left|\phi\right\rangle,\left|\psi\right\rangle):
\left|\phi\right\rangle_A\rightarrow\left|\phi\right\rangle_A+\left|\psi\right\rangle_A$,
respectively.

Any state $\left|\Phi\right\rangle_{ABC}$, in the
$2\times{M}\times{N}$ space for $2\leq M \leq N \leq 2M$, can be
expanded as its adjoint form
$\left|\Phi\right\rangle_{ABC}=\sum_{i=0}^{N-1}\left|\psi_i\right\rangle_{AB}\left|i\right\rangle_C$,
where $\{\left|\psi_i\right\rangle_{AB},i=0,...,N-1\}$ are
linearly independent and $\{\left|i\right\rangle_C,i=0,...,N-1\}$
are a set of computational basis. According to the conclusion of
the lemma 10 in \cite{Kraus}, we know that there is always at
least one product state in $R(\rho^{AB}_{\Phi_{ABC}})$. Suppose
$\sum_{i=0}^{N-1}c_i\left|\psi_i\right\rangle_{AB}$ is of product
form, where the constants $c_i$'s do not equal zero
simultaneously. Let $c_k\neq0,k\in[0,N-1]$. By performing the
operations $O^C_1(\left|k\right\rangle,c_k)\otimes\prod_{i=0,i\neq
k}^{N-1} O^C_2(\left|i\right\rangle,c_i\left|k\right\rangle)$,
$\left|k\right\rangle_C\leftrightarrow\left|N-1\right\rangle_C$
and some ILO's making
$\sum_{i=0}^{N-1}c_i\left|\psi_i\right\rangle_{AB}\rightarrow\left|0,M-1\right\rangle_{AB}$,
one can obtain $\left|\Phi\right\rangle_{ABC}\sim\sum_{i=0,i\neq
k}^{N-1}\left|\psi_i\right\rangle_{AB}\left|i\right\rangle_C+
\sum_{i=0}^{N-1}c_i\left|\psi_i\right\rangle_{AB}\left|k\right\rangle_C
\sim\left|0,M-1,N-1\right\rangle+\sum_{i=0}^{N-2}
\left|\psi^{\prime}_i\right\rangle_{AB}\left|i\right\rangle_C$.
Thus $\{\left|\psi^{\prime}_i\right\rangle_{AB},i=0,...,N-2\}$
remains a set of linearly independent vectors. Let
$\left|\Psi\right\rangle=\sum_{i=0}^{N-2}
\left|\psi^{\prime}_i\right\rangle_{AB}\left|i\right\rangle_C$, we
have
\begin{eqnarray*}
M&=&rank(\rho^B_{\Phi_{ABC}})
=rank(\left|M-1\right\rangle\left\langle M-1\right|+\rho^B_{\Psi_{ABC}})\\
&&\leq rank(\left|M-1\right\rangle\left\langle
M-1\right|)+rank(\rho^B_{\Psi_{ABC}}),
\end{eqnarray*}
so $rank(\rho^B_{\Psi_{ABC}})\geq M-1$. Based on the restriction
of local ranks, we obtain a general equivalence relation such that
$\left|\Phi\right\rangle_{ABC}\sim\left|0,M-1,N-1\right\rangle+
\left|\Psi^{\prime}\right\rangle_{ABC}$,
$\left|\Psi^{\prime}\right\rangle_{ABC}$ lies in $(
\left|\Psi_0\right\rangle_{2\times(M-1) \times(N-1)},
\left|\Psi_1\right\rangle_{2\times M \times (N-1)})$ and
$(\left|\Psi_2\right\rangle_{1\times(M-1) \times(N-1)},
\left|\Psi_3\right\rangle_{1\times M \times (N-1)})$. That is, any
state $\left|\Psi\right\rangle_{2\times M \times N}$ can be
equivalently transformed into one of the four kinds of states by
some ILO's. We are going to simplify this relation so that it is
more efficient. Let $2\leq M \leq N$, without loss of generality.
If $M=N$, the ILO's make
$\left|0,M-1,N-1\right\rangle+\left|\Psi_2\right\rangle\sim
\left|0,M-1,N-1\right\rangle+\left|\Psi_3\right\rangle\sim\left|0,M-1,M-1\right\rangle
+\left|1\right\rangle\sum_{i=0}^{M-2}\left|ii\right\rangle\sim\left|1,M-1,M-1\right\rangle+
(\left|000\right\rangle+\left|1\right\rangle\sum_{i=1}^{M-2}\left|ii\right\rangle)\sim
\left|0,M-1,N-1\right\rangle+\left|\Psi_0\right\rangle$. So we
only calculate
$\left|0,M-1,N-1\right\rangle+\left|\Psi_0\right\rangle$ and
$\left|0,M-1,N-1\right\rangle+\left|\Psi_1\right\rangle$. If
$M<N$, since
$\left|0,M-1,N-1\right\rangle+\left|\Psi_2\right\rangle$ leads to
a class with lower rank and so does
$\left|0,M-1,N-1\right\rangle+\left|\Psi_3\right\rangle$ except
the case $N-1=M$, we do not consider them for true entangled
states. The situation of $N-1=M$ is fully similar to that of $N=M$
as above. Now, we arrive at the following conclusion.

\textit{Lemma 1.} For the classification of true tripartite
entangled states, there exists a general equivalence relation
under SLOCC such that
\begin{eqnarray*}
\left|\Psi\right\rangle_{2\times M \times
N}\sim\left|0,M-1,N-1\right\rangle+\left\{\begin{array}{l}
\left|\Psi_0\right\rangle_{2\times(M-1) \times(N-1)};\\
\left|\Psi_1\right\rangle_{2\times M \times (N-1)}.
\end{array}\right.
\end{eqnarray*}
By extracting further the adjoint product state of the $B$ party
from $\left|\Psi_1\right\rangle$ and using of the above lemma, we
obtain
\begin{eqnarray*}
\left|0,M-1,N-1\right\rangle+\left|\Psi_1\right\rangle_{2\times M
\times(N-1)}\sim\left|0,M-1,N-1\right\rangle\\
+\left|1,M-1\right\rangle\left|\chi\right\rangle+\left\{\begin{array}{l}
\left|\Phi_0\right\rangle_{2\times(M-1) \times(N-1)};\\
\left|\Phi_1\right\rangle_{2\times(M-1) \times(N-2)},
\end{array}\right.
\end{eqnarray*}
where $\left|\chi\right\rangle\equiv\sum_{i=0}^{N-2}
a_i\left|i\right\rangle$ and the arbitrary constants $a_i$'s do
not equal zero simultaneously. By combining the above result with
the lemma, we can write out an united relation of equivalence.

\textit{Corollary 2.} For the classification of true tripartite
entangled states under SLOCC, the following equivalence relation
is
true, \\
$ \left|\Psi\right\rangle_{2\times M \times N}\sim
\left\{\begin{array}{l}
\left|\Omega_0\right\rangle\equiv(a\left|0\right\rangle
+b\left|1\right\rangle)\left|M-1,N-1\right\rangle
\\+\left|\Psi\right\rangle_{2\times(M-1) \times(N-1)},\\
\left|\Omega_1\right\rangle\equiv\left|0,M-1,N-1\right\rangle
\\+\left|1,M-1,N-2\right\rangle+\left|\Psi\right\rangle_{2\times(M-1)\times(N-2)},\\
\left|\Omega_2\right\rangle\equiv\left|\Omega_0\right\rangle
+\left|0,M-1\right\rangle\left|\chi\right\rangle,b\neq0,\\
\left|\Omega_3\right\rangle\equiv\left|\Omega_0\right\rangle
+\left|1,M-1\right\rangle\left|\chi\right\rangle,a\neq0.
\end{array}\right.$\\
The condition $a\neq0$ or $b\neq0$ keeps
$\left|\Omega_2\right\rangle$ and $\left|\Omega_3\right\rangle$
not becoming $\left|\Omega_0\right\rangle$.

Such equivalence relation shows that the lower rank classes of the
entangled states can be used to generate the higher rank classes
of the true entangled states for any $2\times M\times N$ system,
called as "Low-to-High Rank Generating Mode" or LHRGM for short.
So the corollary and the range criterion of the theorem 1 provide
a systematic method to classify all kinds of true tripartite
entangled states in the $2\times M\times N$ system.

First of all, we re-derive the result of true three qubit
entanglements in the present formulation. For the $2\times 2\times
2$ system, from corollary 2,
$\left|\Omega_0\right\rangle_{2\times2\times2}\sim(a\left|0\right\rangle
+b\left|1\right\rangle)\left|11\right\rangle
+\left|000\right\rangle$. It is transformed into
$\left|111\right\rangle +\left|000\right\rangle$ by the ILO
$O^A_2(\left|1\right\rangle,-ab^{-1}\left|0\right\rangle)$ for
$b\neq0$. Hence,
$\left|\Omega_0\right\rangle_{2\times2\times2}\sim\left|GHZ\right\rangle$.
By means of
$\left|0\right\rangle_A\leftrightarrow\left|1\right\rangle_A$, one
can easily see that
$\left|\Omega_2\right\rangle\sim\left|\Omega_3\right\rangle$. So
we only calculate
$\left|\Omega_2\right\rangle_{2\times2\times2}\sim(a\left|0\right\rangle
+b\left|1\right\rangle)\left|11\right\rangle
+(c\left|0\right\rangle
+d\left|1\right\rangle)\left|00\right\rangle+\left|010\right\rangle$.
By performing the same ILO's on
$\left|\Omega_0\right\rangle_{2\times2\times2}$, we can transform
$\left|\Omega_2\right\rangle_{2\times2\times2}$ into
$\left|111\right\rangle +(c^\prime\left|0\right\rangle
+d^\prime\left|1\right\rangle)\left|00\right\rangle+\left|010\right\rangle$.
If $c^\prime\neq0$, we find that
$\left|\Omega_2\right\rangle_{2\times2\times2}\sim\left|GHZ\right\rangle$
by the ILO's $O_2$ acting on each party. On the other hand, if
$c^\prime=0$, one readily gets
$\left|\Omega_2\right\rangle_{2\times2\times2}\sim\left|W\right\rangle$
by $\left|0\right\rangle_A\leftrightarrow\left|1\right\rangle_A$
and $\left|0\right\rangle_B\leftrightarrow\left|1\right\rangle_B$.
$\left|GHZ\right\rangle$ and $\left|W\right\rangle$ are
characterized as two states belonging to the classes $[2,2,2]$ and
$[1,1,1]$, respectively. According to the range criterion of the
theorem 1, they are two inequivalent states of true tripartite
entanglement under SLOCC.

Now we turn to a more complicated system, the $2\times3\times3$
system. It can be easily seen from the corollary that all possible
classes of true entanglement in this system can be written as the
following forms\\
$\left|022\right\rangle+V_A\otimes V_B\otimes V_C\left|\Psi\right\rangle_{2\times2\times2},\\
\left|022\right\rangle+\left|120\right\rangle+V_A\otimes
V_B\otimes V_C\left|\Psi\right\rangle_{2\times2\times2},\\
\left|022\right\rangle+\left|121\right\rangle+\left|000\right\rangle+\left|110\right\rangle,$\\
where $V_A,V_B,V_C$ are $2\times2$ ILO's.

Based on the method of the LHRGM, we can use the lower rank
entangled classes $\left|GHZ\right\rangle$ and
$\left|W\right\rangle$ to generate some entangled classes of the
$2\times3\times3$ system. By using of the explicit forms of
$2\times2$ ILO's, the above expressions can be
equivalently written  as\\
$(a\left|0\right\rangle+b\left|1\right\rangle)\left|22\right\rangle
+\left|000\right\rangle+\left|111\right\rangle, \hspace*{\fill}(I)\\
(a\left|0\right\rangle+b\left|1\right\rangle)\left|22\right\rangle
+\left|001\right\rangle+\left|010\right\rangle+\left|100\right\rangle, \hspace*{\fill}(II)\\
(a\left|0\right\rangle+b\left|1\right\rangle)\left|22\right\rangle+
(c\left|0\right\rangle+d\left|1\right\rangle)\left|2\right\rangle(f\left|0\right\rangle
+g\left|1\right\rangle)+\left|000\right\rangle+\left|111\right\rangle,\hspace*{\fill}(III)\\
(a\left|0\right\rangle+b\left|1\right\rangle)\left|22\right\rangle+
(c\left|0\right\rangle+d\left|1\right\rangle)\left|2\right\rangle(f\left|0\right\rangle
+g\left|1\right\rangle)+\left|001\right\rangle+\left|010\right\rangle+\left|100\right\rangle, \hspace*{\fill}(IV)\\
\left|022\right\rangle+\left|121\right\rangle+\left|000\right\rangle+\left|110\right\rangle.\hspace*{\fill}(V)$\\
Here, $(a,b),(c,d),(f,g)$ are arbitrary constants and two
constants in each bracket cannot equal zero simultaneously. By
classifying the expressions (I), $\cdots$, (V) into all possible
entangled classes under ILO's, one can find out all inequivalent
classes of true entanglement in the $2\times3\times3$ system. The
result is following.

\textit{Theorem 2}. There are six classes of true entangled states
under SLOCC in the $2\times3\times3$ system as following\\
$\left|\Psi_1\right\rangle=\left|000\right\rangle+\left|111\right\rangle
+(\left|0\right\rangle+\left|1\right\rangle)\left|22\right\rangle\in[0,3,3],\\
\left|\Psi_2\right\rangle=\left|010\right\rangle+\left|001\right\rangle
+\left|112\right\rangle+\left|121\right\rangle\in[0,\infty,\infty],\\
\left|\Psi_3\right\rangle=\left|000\right\rangle+\left|111\right\rangle
+\left|022\right\rangle\in[1,\infty,\infty],\\
\left|\Psi_4\right\rangle=\left|100\right\rangle+\left|010\right\rangle
+\left|001\right\rangle+\left|112\right\rangle+\left|121\right\rangle\in[0,1,1],\\
\left|\Psi_5\right\rangle=\left|100\right\rangle+\left|010\right\rangle
+\left|001\right\rangle+\left|022\right\rangle\in[1,\infty,\infty],\\
\left|\Psi_6\right\rangle=\left|100\right\rangle+\left|010\right\rangle
+\left|001\right\rangle+\left|122\right\rangle\in[0,2,2].$

\textit{Proof}. According to the range criterion, we only have to
judge state $\left|\Psi_3\right\rangle$ and
$\left|\Psi_5\right\rangle$. We notice the product states are
$\left|11\right\rangle$ in $R(\rho^{BC}_{\Psi_3})$ and
$\left|00\right\rangle$ in $R(\rho^{BC}_{\Psi_5})$. This implies
that if they are transformable, $\left|0\right\rangle^C_5$ must be
transformed into $\left|1\right\rangle^C_3$. Because the adjoint
state of $\left|0\right\rangle^C_5$ is
$\left|01\right\rangle+\left|10\right\rangle$, and the adjoint
state of $\left|1\right\rangle^C_3$ is $\left|11\right\rangle$, so
there is no ILO between these two states.

Subsequently, we have to prove that the expressions (I), $\cdots$,
(V) can be only and just transformed into the states in the
theorem 2 by the ILO's. First we observe the expression (I). If
$ab=0$, by
$\left|0\right\rangle\leftrightarrow\left|1\right\rangle$ in all
systems and $O^B_1(\left|2\right\rangle,\alpha)$ ( $\alpha$ is
regarded as the possible $a^{-1}$, $b^{-1}$, etc ), it leads to
the state $\left|\Psi_3\right\rangle$. On the other hand, if
$ab\neq0$, the operation
$O^A_1(\left|1\right\rangle,\alpha^{-1})\otimes
O^B_1(\left|1\right\rangle,\alpha)$ makes the expression (I) into
the state $\left|\Psi_1\right\rangle$. Second, the expression
(II), if $b=0$, becomes $\left|\Psi_5\right\rangle$. For $b\neq0$,
by means of
$O^A_2(\left|1\right\rangle,\alpha\left|0\right\rangle)\otimes
O^C_2(\left|1\right\rangle,-\alpha\left|0\right\rangle)$, we can
transform (II) into the state $\left|\Psi_6\right\rangle$. The
ILO's $O_2$ and $O_1$ acting on each party produce that the
repression (IV) $\sim
\left|122\right\rangle+\left|02\right\rangle(f^{\prime}\left|0\right\rangle
+g^{\prime}\left|1\right\rangle)+
\left|001\right\rangle+\left|010\right\rangle+\left|100\right\rangle$
for $b\neq0$ and $\sim \left|022\right\rangle+
\left|121\right\rangle+
\left|001\right\rangle+\left|010\right\rangle+\left|100\right\rangle$
for $b=0$. Furthermore, by the ILO's of $O^C_2$,
$\left|0\right\rangle_B\leftrightarrow\left|1\right\rangle_B$,
$\left|0\right\rangle_C\leftrightarrow\left|1\right\rangle_C$ for
$b\neq0$ and of
$\left|1\right\rangle_B\leftrightarrow\left|2\right\rangle_B$,
$\left|1\right\rangle_C\leftrightarrow\left|0\right\rangle_C$,
$\left|1\right\rangle_A\leftrightarrow\left|0\right\rangle_A$ for
$b=0$, we establish the relation the repression (IV) $\sim
\left|\Psi_6\right\rangle$ for $b\neq0$ and $\sim
\left|\Psi_4\right\rangle$ for $b=0$. The expression (V) leads to
state $\left|\Psi_2\right\rangle$ by
$\left|0\right\rangle_B\leftrightarrow\left|2\right\rangle_B$ and
$\left|1\right\rangle\leftrightarrow\left|0\right\rangle$ in all
parties. Finally, we have checked that the expression (III) is
transformed into the expressions of (I) and (II) under some ILO's.
Q.E.D.

In the $2\times M\times 2M$ system, since the dimension of the
Hilbert space of the $AB$ part is equal to that of the $C$ part in
the Schmidt decomposition, all pure states of this system are
transformed into an unique equivalent class of true entanglement
$\left|\Upsilon_0\right\rangle\equiv\left|0\right\rangle\sum_{i=0}^{M-1}\left|ii\right\rangle
+\left|1\right\rangle\sum_{i=0}^{M-1}\left|i,i+M\right\rangle\in[0
,0,\infty]$ under the ILO's. By using of this result and the
method of LHRGM, we can construct the equivalent classes of true
entanglement in a general type of the $2\times M\times N$ system.

\textit{Theorem 3}. There are two classes of states under SLOCC in
any $2\times(M+1)\times(2M+1)$ system ($M\geq1$),\\
$\left|\Upsilon_1\right\rangle\equiv\left|0,M,2M\right\rangle+\left|\Upsilon_0\right\rangle
\in[0,1,\infty]$;\\
$\left|\Upsilon_2\right\rangle\equiv\left|0,M,2M\right\rangle+\left|1,M,M-1\right\rangle
+\left|\Upsilon_0\right\rangle\in[0,0,\infty]$.

\textit{Proof.} Following the LHRGM, from corollary 2, we can read
out $\left|\Omega_0\right\rangle\sim(a\left|0\right\rangle
+b\left|1\right\rangle)\left|M,2M\right\rangle
+\left|\Upsilon_0\right\rangle$. If ab=0,
$\left|\Omega_0\right\rangle$ just is
$\left|\Upsilon_1\right\rangle$. For the case of $ab\neq0$, we
perform the ILO $U_1=
O^A_2(\left|0\right\rangle,\alpha\left|1\right\rangle)\otimes\prod_{i=0}^{M-1}
O^C_2(\left|i+M\right\rangle,-\alpha\left|i\right\rangle)$ on
$\left|\Omega_0\right\rangle$ to be reduced to
$\left|\Upsilon_1\right\rangle$.  In the below proof, the
invariance of $\left|\Upsilon_0\right\rangle$ under two ILO's is
very useful. They are given by ILO \textbf{I}:
$\left|0\right\rangle_A\leftrightarrow\left|1\right\rangle_A$,
$\left|i\right\rangle_C\leftrightarrow\left|i+M\right\rangle_C,$
$i=0,...,M-1,$ and ILO \textbf{II}:
$\left|i\right\rangle_B\leftrightarrow\left|j\right\rangle_B$,
$\left|i\right\rangle_C\leftrightarrow\left|j\right\rangle_C$,
$\left|i+M\right\rangle_C\leftrightarrow\left|j+M\right\rangle_C,i,j\in\{0,...,M-1\}$.
By the ILO \textbf{I}'s invariance of
$\left|\Upsilon_0\right\rangle$, it is obvious that
$\left|\Omega_2\right\rangle\sim\left|\Omega_3\right\rangle
\sim(a\left|0\right\rangle+b\left|1\right\rangle)\left|M,2M\right\rangle+
\left|1,M\right\rangle\left|\chi\right\rangle+\left|\Upsilon_0\right\rangle$.
By using the ILO's $U_1$, $U_2=\prod_{i=0}^{M-1}
O^B_2(\left|i\right\rangle,-a_i\left|M\right\rangle)\otimes
O^C_2(\left|2M\right\rangle,\sum_{i=0}^{M-1}a_i\left|i\right\rangle)$,
and the modified $U_2$ to act on $\left|\Omega_3\right\rangle$
sequently, we can transform $\left|\Omega_3\right\rangle$ into
$\left|\Upsilon_2\right\rangle$ by means of the ILO \textbf{II}'s
invariance of $\left|\Upsilon_0\right\rangle$.

Finally we determine the $\left|\Omega_1\right\rangle$'s family by
induction. First of all, let us notice that
$\left|\Omega_1\right\rangle_{2\times (M+2)\times
(2M+3)}\sim\left|0,M+1,2M+2\right\rangle+\left|1,M+1,2M+1\right\rangle
+\left|\Upsilon_i\right\rangle_{2\times (M+1)\times (2M+1)}$ leads
to the states $\left|\Upsilon_i\right\rangle_{2\times (M+2)\times
(2M+3)}$, by both
$\left|M+1\right\rangle_B\leftrightarrow\left|M\right\rangle_B$
and
$\left|2M+2\right\rangle_C\leftrightarrow\left|2M\right\rangle_C$.
This iterated relation implies that if there are only two classes
in $\left|\Psi\right\rangle_{2\times (M+1)\times (2M+1)}$,
$\left|\Omega_1\right\rangle_{2\times (M+2)\times (2M+3)}$ must
enter into two classes of states in
$\left|\Psi\right\rangle_{2\times(M+2)\times(2M+3)}$. For the case
$M=1$, the conclusion of \cite{Miyake2} told us that indeed there
exist only two classes of true entangled states in the $2\times
2\times 3$ system. Hence, there exist only two inequivalent
classes of $\left|\Psi\right\rangle_{2\times (M+1)\times (2M+1)}$
under SLOCC, and $\left|\Omega_1\right\rangle$ must be classified
into such two classes $\left|\Upsilon_1\right\rangle$ and
$\left|\Upsilon_2\right\rangle$. Q.E.D.

Subsequently, we can use the entangled classes in the theorem 3 to
generate those in another type of systems in the LHRGM way. The
structure of the entangled classes in this type of system is
different from that in theorem 3. We give them here.

\textit{Theorem 4}. In any $2\times(M+2)\times(2M+2)$
system($M\geq2$), there are six classes of true entangled states
under SLOCC\\
$ \left|\Theta_0\right\rangle\equiv\left|1,M+1,2M+1\right\rangle+
\left|\Upsilon_1\right\rangle\in[0,2,\infty];\\
\left|\Theta_1\right\rangle\equiv\left|0,M+1,2M+1\right\rangle+
\left|\Upsilon_1\right\rangle\in[0,\infty,\infty];\\
\left|\Theta_2\right\rangle\equiv\left|1,M+1,2M+1\right\rangle+
\left|\Upsilon_2\right\rangle\in[0,1,\infty];\\
\left|\Theta_3\right\rangle\equiv\left|0,M+1,2M+1\right\rangle+
\left|1,M+1,2M\right\rangle+\left|\Upsilon_1\right\rangle\in[0,1,\infty];\\
\left|\Theta_4\right\rangle\equiv\left|0,M+1,2M+1\right\rangle+
\left|1,M+1,0\right\rangle+\left|\Upsilon_2\right\rangle\in[0,0,\infty];\\
\left|\Theta_5\right\rangle\equiv\left|0,M+1,2M+1\right\rangle+
\left|1,M+1,2M\right\rangle+\left|\Upsilon_2\right\rangle\in[0,0,\infty].$

\textit{Proof.} According to corollary 2 and the rule of LHRGM,
first, we write out the states to be dealt with\\
$\left|\Omega_0\right\rangle\sim(a\left|0\right\rangle+b\left|1\right\rangle)\left|M+1,2M+1\right\rangle
+\left|\Upsilon_j\right\rangle,$\\
$\left|\Omega_2\right\rangle\sim\left|\Omega_0\right\rangle
+\left|0,M+1\right\rangle\sum_{i=0}^{2M}a_i\left|i\right\rangle,$\\
$\left|\Omega_3\right\rangle\sim\left|\Omega_0\right\rangle
+\left|1,M+1\right\rangle\sum_{i=0}^{2M}a_i\left|i\right\rangle,$\\
where $j=1,2$. Now, we have to prove that the above states must be
transformed into six kinds of states in the theorem 4 by some
ILO's. These ILO's are composed of those similar to $U_1$, $U_2$
appeared in the proof of the theorem 3, and the ILO's of state
exchanges in each party. By means of these ILO's, we have proven
the following conclusions. For the case of generation in the
direction of the class $\left|\Upsilon_1\right\rangle$, it can be
easily proved that, by ILO's
$\widetilde{U_1}=O^B_2(\left|M\right\rangle,-a_{2M}\left|M+1\right\rangle)$,
and $
\left|M+1\right\rangle_B\leftrightarrow\left|M\right\rangle_B,
\left|2M+1\right\rangle_C\leftrightarrow\left|2M\right\rangle_C$,
we have
$\left|\Omega_2\right\rangle\sim\left|\Omega_0\right\rangle$. By
ILO $
\widetilde{U_2}={O^A_2(\left|1\right\rangle,\alpha\left|0\right\rangle)}\otimes
{\prod_{i=0}^{M-1}O^C_2(\left|i\right\rangle,-\alpha\left|i+M\right\rangle)}$,
$\left|\Omega_0\right\rangle$ is transferred to
$\left|\Theta_0\right\rangle$ for $b\neq0$, and
$\left|\Theta_1\right\rangle$ for $b=0$ respectively. Meanwhile,
by
$\widetilde{U_3}=O^A_2(\left|1\right\rangle,\alpha\left|0\right\rangle)\otimes
O^C_2(\left|2M+1\right\rangle,-\sum_{i=0}^{2M}a_i\left|i\right\rangle)\otimes
O^B_2(\left|M\right\rangle,-\alpha\left|M+1\right\rangle)$, and $
\left|M+1\right\rangle_B\leftrightarrow\left|M\right\rangle_B,
\left|2M+1\right\rangle_C\leftrightarrow\left|2M\right\rangle_C$,
we obtain
$\left|\Omega_3\right\rangle\sim\left|\Omega_0\right\rangle$ for
$b\neq0$. For $b=0$,
$\left|\Omega_3\right\rangle\sim\left|\Omega_0\right\rangle$ if
there is at least one non-vanishing $a_i,i=0,1,...,M-1$, and
$\left|\Omega_3\right\rangle\sim\left|\Theta_3\right\rangle$ if
all $a_i$'s equal zero, $i=0,1,...,M-1$.

On the other hand, we consider the generation in the direction of
the class $\left|\Upsilon_2\right\rangle$. It is useful that the
invariance of $\left|\Upsilon_2\right\rangle$ under the ILO
\textbf{III}:
$\left|0\right\rangle_A\leftrightarrow\left|1\right\rangle_A,
\left|M\right\rangle_B\leftrightarrow\left|M-1\right\rangle_B,
\left|2M\right\rangle_C\leftrightarrow\left|2M-1\right\rangle_C$
and
$\left|i\right\rangle_C\leftrightarrow\left|i+M\right\rangle_C,i=0,...,M-2$.
By virtue of this invariance and the ILO's, we find that
$\left|\Omega_0\right\rangle\sim\left|\Theta_2\right\rangle$.
Furthermore,
$\left|\Omega_2\right\rangle\sim\left|\Omega_3\right\rangle\sim\left|\Theta_4\right\rangle$
if there is at least one non-vanishing $a_i,i=0,1,...,M-2$, and
$\left|\Omega_2\right\rangle\sim\left|\Omega_3\right\rangle\sim\left|\Theta_5\right\rangle$
if all $a_i$'s equal zero, $i=0,1,...,M-2$.

Similarly, one can calculate $\left|\Omega_1\right\rangle$ by
induction. By calculation of $\left|\Omega_0\right\rangle$,
$\left|\Omega_2\right\rangle$ and $\left|\Omega_3\right\rangle$
above, the classes in $\left|\Psi\right\rangle_{2\times3\times4}$
are found such that\\
$\left|\Theta_0\right\rangle_{2\times3\times4}\sim
\left|123\right\rangle+\left|012\right\rangle+\left|000\right\rangle+\left|101\right\rangle,\\
\left|\Theta_1\right\rangle_{2\times3\times4}\sim
\left|023\right\rangle+\left|012\right\rangle+\left|000\right\rangle+\left|101\right\rangle,\\
\left|\Theta_2\right\rangle_{2\times3\times4}\sim
\left|123\right\rangle+\left|012\right\rangle+\left|110\right\rangle
+\left|000\right\rangle+\left|101\right\rangle,\\
\left|\Theta_3\right\rangle_{2\times3\times4}\sim
\left|023\right\rangle+\left|122\right\rangle
+\left|012\right\rangle+\left|000\right\rangle+\left|101\right\rangle,\\
\left|\Theta_5\right\rangle_{2\times3\times4}\sim
\left|023\right\rangle+\left|122\right\rangle
+\left|012\right\rangle+\left|110\right\rangle+\left|000\right\rangle+\left|101\right\rangle.$\\
Notice that the class $\left|\Theta_4\right\rangle$ disappears
here, for the coefficients $a_i,i=0,...,M-2$ always equal zero in
the above derivation of $\left|\Theta_4\right\rangle$ and
$\left|\Theta_5\right\rangle$. Then
$\left|\Omega_1\right\rangle_{2\times3\times4}\sim
\left|023\right\rangle+\left|122\right\rangle+\left|\Psi\right\rangle_{2\times2\times2}$.
For the case of
$\left|\Psi\right\rangle_{2\times2\times2}\sim\left|GHZ\right\rangle$,
we have
$\left|\Omega_1\right\rangle\sim\left|023\right\rangle+\left|122\right\rangle+
\left|000\right\rangle+\left|111\right\rangle$. By
$\left|3\right\rangle_C\leftrightarrow\left|1\right\rangle_C$ and
$\left|2\right\rangle_B\leftrightarrow\left|1\right\rangle_B$, one
obtains $\left|\Omega_1\right\rangle\sim\left|123\right\rangle
+\left|\Psi\right\rangle_{2\times2\times3}\sim\left|\Omega_0\right\rangle$.
On the other hand, if
$\left|\Psi\right\rangle_{2\times2\times2}\sim\left|W\right\rangle$,
then
$\left|\Omega_1\right\rangle\sim\left|023\right\rangle+\left|122\right\rangle+
\left|001\right\rangle+\left|010\right\rangle+\left|100\right\rangle$.
By the operations
$\left|3\right\rangle_C\leftrightarrow\left|1\right\rangle_C$ and
$\left|2\right\rangle_B\leftrightarrow\left|0\right\rangle_B$, we
get
$\left|\Omega_1\right\rangle\sim\left|001\right\rangle+\left|102\right\rangle+
\left|023\right\rangle+\left|010\right\rangle+\left|120\right\rangle
\sim\left|023\right\rangle+\left|120\right\rangle+\left|\Psi\right\rangle_{2\times2\times3}
\sim\left|\Omega_3\right\rangle$. So there are five classes of
entanglement in the $2\times3\times4$ system. Next, one should
write out all classes in
$\left|\Psi\right\rangle_{2\times4\times6}$, which is really the
first step of the induction. Following the above technique we
obtain
$\left|\Theta_i\right\rangle_{2\times4\times6},i=0,1,2,3,4,5$.
Then we calculate
$\left|\Omega_1\right\rangle_{2\times4\times6}\sim\left|035\right\rangle+
\left|134\right\rangle+\left|\Psi\right\rangle_{2\times3\times4}$,
where $\left|\Psi\right\rangle_{2\times3\times4}\sim
\left|\Theta_i\right\rangle_{2\times3\times4},i=0,1,2,3,5.$ By
$\left|5\right\rangle_C\leftrightarrow\left|3\right\rangle_C$ and
$\left|3\right\rangle_B\leftrightarrow\left|2\right\rangle_B$, for
the case of
$\left|\Theta_i\right\rangle_{2\times3\times4},i=0,1,2$,
$\left|\Omega_1\right\rangle\sim\left|135\right\rangle+\left|\Psi\right\rangle_{2\times3\times5}
\sim\left|\Omega_0\right\rangle$; for the case of
$\left|\Theta_i\right\rangle_{2\times3\times4},i=3,5$,
$\left|\Omega_1\right\rangle\sim\left|035\right\rangle+\left|132\right\rangle+
\left|\Psi\right\rangle_{2\times3\times5}\sim\left|\Omega_3\right\rangle$.
So there are six classes of entanglement in the $2\times4\times6$
system, i.e.,
$\left|\Theta_i\right\rangle_{2\times4\times6},i=0,1,2,3,4,5$.
Similar to the first step, one can continue with the deduction.
That is, by
$\left|2M+3\right\rangle_C\leftrightarrow\left|2M+1\right\rangle_C$
and
$\left|M+2\right\rangle_B\leftrightarrow\left|M+1\right\rangle_B$,
we always get
$\left|\Omega_1\right\rangle_{2\times(M+3)\times(2M+4)}\sim
\left|0,M+2,2M+3\right\rangle+\left|1,M+2,2M+2\right\rangle+
\left|\Psi\right\rangle_{2\times(M+2)\times(2M+2)}\sim\left|\Omega_i\right\rangle,i=0,2,3.$
Thus, all classes of $\left|\Omega_1\right\rangle$ belong to
$\left|\Theta_i\right\rangle,i=0,1,2,3,4,5.$

Finally we should check that these classes of states in
$\left|\Psi\right\rangle_{2\times(M+2)\times(2M+2)}$ are
inequivalent. According to the range criterion, we need only to
discuss the relations between $\left|\Theta_2\right\rangle$ and
$\left|\Theta_3\right\rangle$, and between
$\left|\Theta_4\right\rangle$ and $\left|\Theta_5\right\rangle$.
The proof of the former case is similar to that in theorem 2.
Since the requirement of
$\left|2M\right\rangle^C_{\Theta3}\mapsto\left|2M+1\right\rangle^C_{\Theta2}$
always leads to an entangled adjoint state of
$\left|2M+1\right\rangle^C_{\Theta2}$,
$\left|\Theta_2\right\rangle$ and $\left|\Theta_3\right\rangle$
are inequivalent. The proof of the latter case is some difficult.
We can prove it by the reduction to absurdity. First, we suppose
that the AB system is in the adjoint state and there exist the
possible ILO's $V_A$ and $V_B$ to make
$\left|\Theta_4\right\rangle\mapsto\left|\Theta_5\right\rangle$.
By theorem 1, $V_A\otimes V_B[R(\rho^{AB}_{\Theta4})]\in
R(\rho^{AB}_{\Theta5})$. This relation produces a set of equations
satisfied by the matrix elements of $V_A$ and $V_B$. Carefully
analyzing this set of equations, we have found that there exist
only some singular solutions of $V_A$ and $V_B$, i.e. Det$[V_B]=0$
or Det$[V_A]=0$. So there exists no ILO making
$\left|\Theta_4\right\rangle$ and $\left|\Theta_5\right\rangle$
equivalent. We have detailedly proven this fact in the Appendix.
Q.E.D.

To summarize, by using of the range criterion and the method of
LHRGM developed by us here, we have finished the construction of
true entangled classes of some types of the $2\times M\times N$
system, which are with finite kinds of states. For the system with
higher dimensions, one can also find out the classification by the
techniques in this paper \cite{notation2}. However, our method can
be also applied to classify the entangled system with infinite
kinds of states, which is also a puzzling issue in quantum
information theory. Our results are helpful to classify and
construct the inequivalent classes of entangled states in
many-qubit system.

The work was partly supported by the NNSF of China Grant
No.90503009 and 973 Program Grant No.2005CB724508.

\begin{center}
{\bf APPENDIX: THE INEQUIVALENCE OF $\left|\Theta_4\right\rangle$
AND $\left|\Theta_5\right\rangle$}
\end{center}
Suppose that the AB system is in the adjoint
state and the possible ILO's are taken as
\[V_A^{2\times2}=\left(\begin{array}{cc}
w & x \\
y & z
\end{array}\right),V_{B}^{(M+2)\times(M+2)}=[a_{ij}],i,j=0,...,M+1.\]
Thus, these ILO's make
$\left|\Theta_4\right\rangle\mapsto\left|\Theta_5\right\rangle$,
if det$[V_A]$det$[V_B]\neq0$ is satisfied. According to theorem 1,
we must have
\begin{equation*}
V_A\otimes V_B[R(\rho^{AB}_{\Theta4})]\in
R(\rho^{AB}_{\Theta5}),\eqno(A1)
\end{equation*}
which yields that ($M>2$)\\ $ V_A\otimes
V_B[c^\prime_0\left|0,M\right\rangle+\left|0\right\rangle\sum_{i=1}^{M-2}c^\prime_i\left|i\right\rangle
+c^\prime_{M-1}\left|0,M+1\right\rangle+\left|1\right\rangle\nonumber\\
\sum_{i=M}^{2M-1}c^\prime_i\left|i\right\rangle
+c^\prime_{2M}(\left|1,M+1\right\rangle+\left|0,0\right\rangle)
+c^\prime_{2M+1}(\left|1,M\right\rangle+\left|0\right\rangle\nonumber\\
\left|M-1\right\rangle)]
\in\sum_{i=0}^{M-2}c_i\left|0,i\right\rangle
+c_{M-1}\left|0,M+1\right\rangle+
\sum_{i=M}^{2M-1}c_i\left|1,i\right\rangle\nonumber\\
+c_{2M}(\left|1,M+1\right\rangle+\left|0,M\right\rangle)
+c_{2M+1}(\left|1,M\right\rangle+\left|0,M-1\right\rangle),
$
the coefficients $c^\prime_i,c_i$'s, i=0,...,2M+1 are arbitrarily
decided by theorem 1. Let only one $c^\prime_i$ be nonzero, e.g.,
$c^\prime_{M-1}\neq0$ and $c^\prime_i=0,i\neq M-1$. The above
expression then implies
\begin{equation*}
ya_{_{M+1,M+1}}=wa_{_{M,M+1}},ya_{_{M,M+1}}=wa_{_{M-1,M+1}}.
\eqno(A2)
\end{equation*}
Notice we have deserted the trivial results that can't bring those
similar to the above relations between the entries of $V_A$ and
$V_B$ without $c_i$'s. We continue in the same vein, that is, to
choose the uniquely nonzero $c^\prime_i$ in turn and obtain a
sequence of relations such that
\begin{equation*}
\left\{\begin{array}{cc}
ya_{_{M+1,i}}=wa_{_{M,i}},i=1,...,M-2,M,M+1,\\
ya_{_{M,i}}=wa_{_{M-1,i}},i=1,...,M-2,M,M+1,\\
za_{_{M+1,i}}=xa_{_{M,i}},i=0,...,M-1,\\
za_{_{M,i}}=xa_{_{M-1,i}},i=0,...,M-1,\\
za_{_{M+1,M}}+ya_{_{M+1,M-1}}=xa_{_{M,M}}+wa_{_{M,M-1}},\\
za_{_{M,M}}+ya_{_{M,M-1}}=xa_{_{M-1,M}}+wa_{_{M-1,M-1}}.
\end{array}\right.   \eqno(A3)
\end{equation*}
If $wxyz\neq0$, it must be that
$a_{_{M-1,i}}=a_{_{M,i}}=a_{_{M+1,i}}=0,i=1,...,M-2,$ since
$wz\neq xy$. By substituting the former four expressions of (A3)
into the last two expressions of it, on the other hand, one can
get $a_{_{M,M}}/y=a_{_{M,M-1}}/z, a_{_{M,M}}/w=a_{_{M,M-1}}/x$.
This implies that $a_{_{M,M}}=a_{_{M,M-1}}=0$, which leads to
$a_{_{M-1,i}}=a_{_{M,i}}=a_{_{M+1,i}}=0,i=1,...,M.$ This
conclusion is just det$[V_B]=0$. So there must be at least one
equaling zero in $\{w,x,y,z\}.$ If $x=0,wz\neq0$ (or
$y=0,wz\neq0$,etc), it will lead to
$a_{_{M,i}}=a_{_{M+1,i}}=0,i=0,...,M,$ again det$[V_B]=0$. So
there is no ILO between $\left|\Theta_4\right\rangle$ and
$\left|\Theta_5\right\rangle.$ Similarly, for the case of $M=2$,
the same conclusion can be obtained. Q.E.D.


\begin{references}
\bibitem{notation1} Otherwise it means $\sum_i
a_i\left|i\right\rangle=k\sum_i b_i\left|i\right\rangle$, where
$k$ is a proportional constant. Because $\{\left|i\right\rangle\}$
are linearly independent, so the above equality requires
$a_i=kb_i$.
\end{references}
\end{document}